\documentclass[conference]{IEEEtran}

\usepackage{subfigure}
\usepackage[noadjust]{cite}

\ifCLASSINFOpdf
   \usepackage[pdftex]{graphicx}
\else
  \usepackage[dvips]{graphicx}
\fi

\usepackage[cmex10]{amsmath}
\usepackage{fixltx2e}

\usepackage{float}
\usepackage{amsmath}
\begin{document}
%
\title{Construction of Adaptive Short LDPC Codes for Distributed Transmit Beamforming}



\author{\IEEEauthorblockN{Ismail Shakeel}
\IEEEauthorblockA{Defence Science and Technology Group, \\
	Edinburgh, South Australia. \\
    email: \textit{Ismail.Shakeel@dst.defence.gov.au}} \\
    \and
\IEEEauthorblockN{Ishtiaq Ahmad} 
\IEEEauthorblockA{Institute for Telecommunications Research,\\
    University of South Australia. \\
	email: \textit{Ishtiaq.Ahmad@unisa.edu.au}}\\
	    \and
\IEEEauthorblockN{Hajime Suzuki} 
\IEEEauthorblockA{Data61,\\
	CSIRO, Australia. \\
	email: \textit{Hajime.Suzuki@csiro.au} }\\
}


\maketitle

\begin{abstract}
One of the challenges often faced with wireless communication systems is its limited range and data-rate. Distributed Transmit Beamforming (DTB) techniques are being developed to address these two issues to provide reliable connectivity
from power-limited distributed users. This paper proposes an adaptive Low Density Parity Check (LDPC) coding scheme for the DTB system.  The proposed scheme constructs powerful LDPC codes with varying code-rates and block-lengths. This feature of the proposed scheme allows the DTB system to optimise its system resources, improve throughput and communicate reliably under large variation of different channel environments. The performance of some of the codes constructed using the proposed scheme is evaluated and compared with the uncoded and other coded-DTB systems.  The results obtained show large gains over the compared systems. The results also show that coding applied to the DTB system drastically reduces the minimum number of distributed transmit nodes required to achieve a target error-rate with the same energy per information bit to noise power spectral density ($E_{b}/N_{0}$).

\end{abstract}


\IEEEpeerreviewmaketitle

\hfill
\section{Introduction}

Distributed Transmit Beamforming (DTB) is a cooperative wireless communication technique that uses multiple transmitters to send a common message to a distant receiver in such a way that their signals constructively combine at the destination receiver \cite{mudumbai_09}.  DTB promises several benefits, including increased power at destination, increased data-rate and increased communication range. In addition, it also improves transmission security and reduces interference to unintended receivers. Similar to many other wireless communication systems, DTB-based systems can be severely affected by the dynamic time-varying nature of the communication channel, resulting in a significant reduction of the system's overall throughput and transmission reliability.  
To cope with this issue, many modern communication systems use adaptive coding (and modulation) schemes that vary code-rates (and modulation) based on the channel condition.  Systems that use these schemes usually have a very small number of useful coding options available, limiting the range of channel conditions (and environments) they can effectively operate under.   Normally, these systems generate a family of codes using a mother code and selectively use puncturing or shortening to obtain several codes of different block-lengths and code-rates.  This approach does not necessarily generate good codes as the mother code is optimally designed for its original rate and length and subsequent codes generated (with different rates and lengths) can exhibit a wider gap from the Shannon capacity limit than the gap observed for the mother code \cite{nguyen_13}.  

The performance of an error-correction code is dependent on both the code-rate $R$ and block-length $N$.  In general, the code performance improves with a decrease in code-rate.  Similarly, for the continuous unconstrained Additive White Gaussian Noise (AWGN) channel, the code performance improves as the block-length $N$ is increased \cite{dolinar_98}.  However, this relationship between block-length $N$ and performance is not always maintained for all types of channels.  For certain types of interference channels, short codes perform more effectively than long codes.  Furthermore, the desired block-length is also dependent on the application used.  For example, short codes are preferred for voice applications where low latency is required. 
 
This paper proposes an adaptive coding scheme and investigates its performance for the DTB system in the Independent Rayleigh fading channel.  This scheme describes a framework that could be used to construct powerful LDPC codes dynamically in real-time using the parameter pair ($N$, $R$) to meet a set of target performance requirements.  It is assumed that the code parameters ($N$, $R$) are chosen adaptively based on the estimated channel condition to maximise the system performance or simply chosen manually by the user.  The proposed scheme can also be used to enhance transmission security by enabling the transmitter and receiver to generate random error correcting codes by sharing only the code parameter pair ($N$, $R$) and a shared key.  
LDPC codes with large block-lengths are shown to approach the  Shannon  limit \cite{chung_01}. \hspace{0.1cm}  However, larger codes require more memory and computational power and increase implementation complexity and the encoding/decoding latency of the system.  Therefore, this paper will focus mainly on the construction and performance of short codes. The constructed codes have a quasi-cyclic structure that allows the design and implementation of simple encoder/decoder pairs. 

The remainder of this paper is organised as follows. Section~\ref{section_DTB} briefly describes and investigates performance of the uncoded DTB system.  Section \ref{section_ldpc} provides a short introduction to protograph-based LDPC codes and then presents the proposed code construction scheme.  This section also presents performance results for the single-input single-output (SISO) system employing some of the codes constructed using the proposed scheme. Performance of the LDPC coded-DTB systems are investigated and compared with other DTB systems in Section  \ref{section_codedDTB}.  Finally, concluding remarks are presented in Section \ref{section_conclu}.

\hfill
\section{Distributed Transmit Beamforming System}
\label{section_DTB}
\begin{figure}[h!]
        \centering\small
        \includegraphics[width=.49\textwidth]{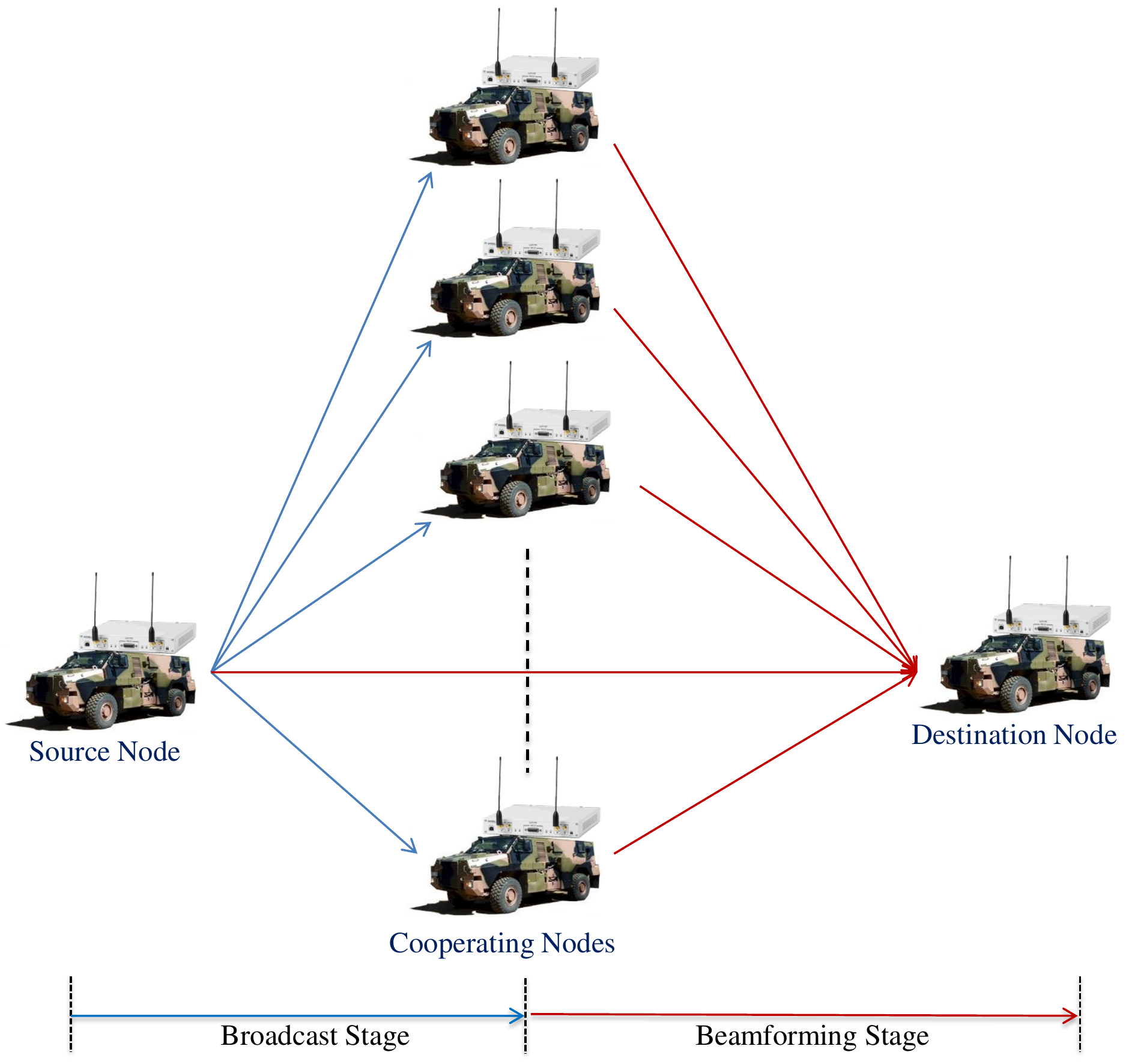}
        \caption{Standard DTB system}
        \label{dtb_system}
\end{figure}

The block diagram of a standard Distributed Transmit Beamforming (DTB) system is depicted in Figure \ref{dtb_system}.  
The DTB system operates in two stages, namely, the broadcast stage and the beamforming stage.  During the broadcast stage, the source node broadcasts its message to the nearby (cooperative) nodes.  Each node can be a source node as well as a cooperative node to other source nodes.  In this paper, we assume the cooperative nodes are in close proximity to the source and the channel between them is error free.  Further, this paper will only focus on the beamforming stage and the analysis presented will not take into account the energy dissipated in the broadcast stage.  During the beamforming stage, the cooperating nodes transmit the message (received from the source node) such that their signals add up coherently at the destination node.  Ideal beamforming with $M$ cooperating nodes can lead to a $M^{2}$-fold increase in received signal to noise power ratio (SNR) at the destination node.  This also translates to an $M$-fold gain in the energy per information bit to noise power spectral density ($E_{b}/N_{0}$) over the non-cooperative SISO system, assuming the total transmit power used by both these systems are kept the same.  One of the key difficulties in transmit beamforming is the synchronisation (phase, time, frequency) of all the individual cooperating signals at the destination.  The performance investigations carried out in this paper assume ideal synchronisation. In addition, we also assume that all the cooperating nodes are at the same distance from the destination node and are of equal transmit power.  

The total received signal,  $\textit{r}$,  at the destination node can be expressed as \cite{tushar_12},
\begin{equation}\label{systemEq}
\textit{r} = \textit{m}\sum_{i=1}^{M} a_{i}e^{j\phi_{i}} + \textit{n}
\end{equation} 
where  $\textit{m}$ is the common message transmitted by each of the cooperating $M$ nodes, $\textit{n}$ is the Additive White Gaussian Noise (AWGN) with mean 0 and variance $\sigma^2_{\textit{n}}$, $\phi_{i}$ is the phase of the transmitted signal from transmitter $i$ and $a_{i}$ denotes the channel from the  cooperating transmitters to the destination.  We assume a time-varying flat fading channel that stays constant during the channel symbol period but varies from symbol to symbol.  We assume a Rayleigh fading channel, $a_{i}$, given by
\begin{equation}\label{fadingEq}
a_{i} = |a_{i}|e^{j\psi_{i}}, 
\end{equation} 
such that $a_{i},  i = 1, ..., M$, are independent and identically distributed (i.i.d.) random variables following $a_{i} \sim \mathcal{C} \mathcal{N} (0, 1)$.    

\begin{figure}[h!]
	\centering\small
	\includegraphics[width=.53\textwidth]{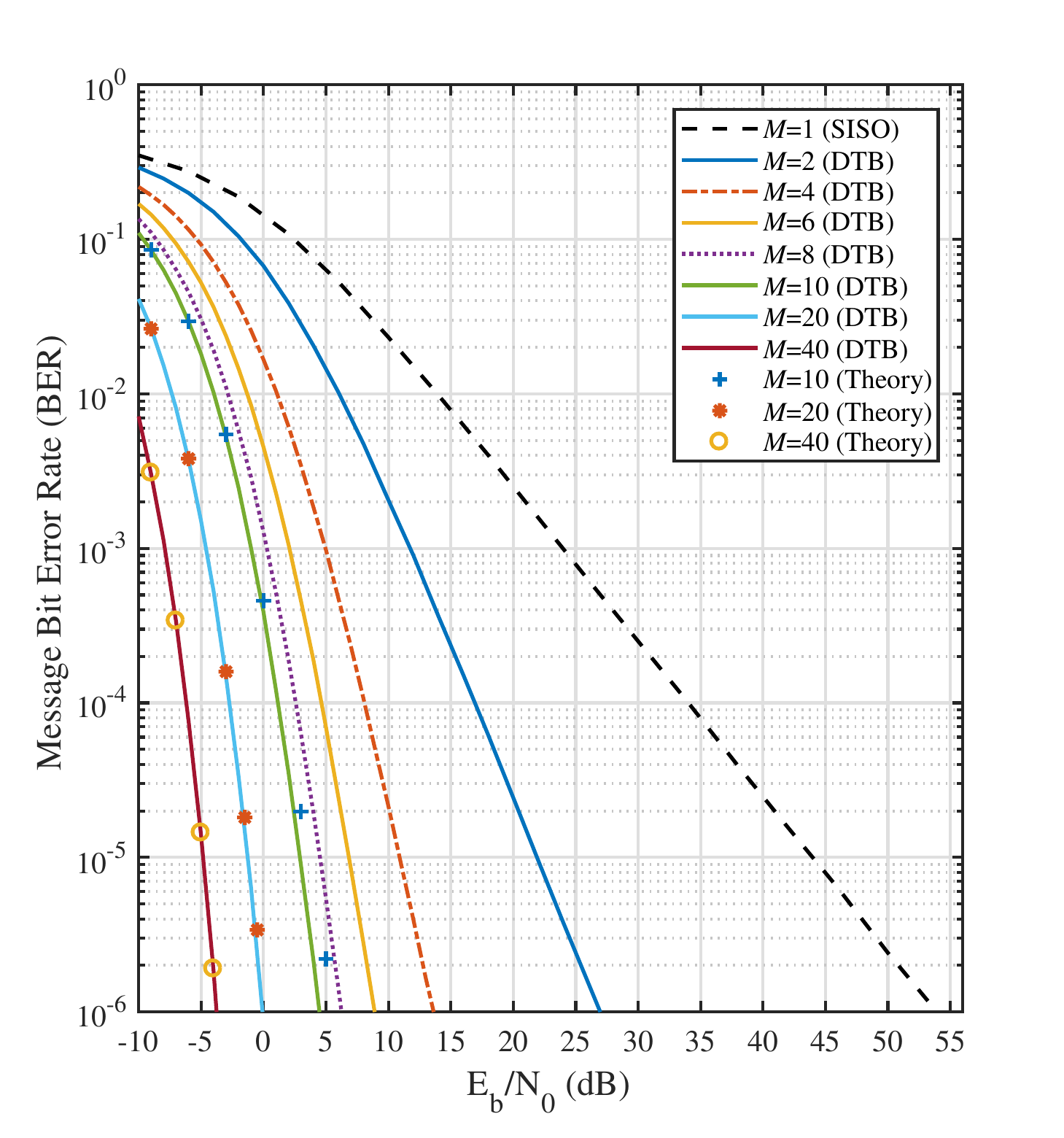}
	\caption{Performance of uncoded DTB systems in Independent Rayleigh fading}
	\label{uncoded}
\end{figure}

\begin{figure}[h!]
	\centering\small
	\includegraphics[width=.53\textwidth]{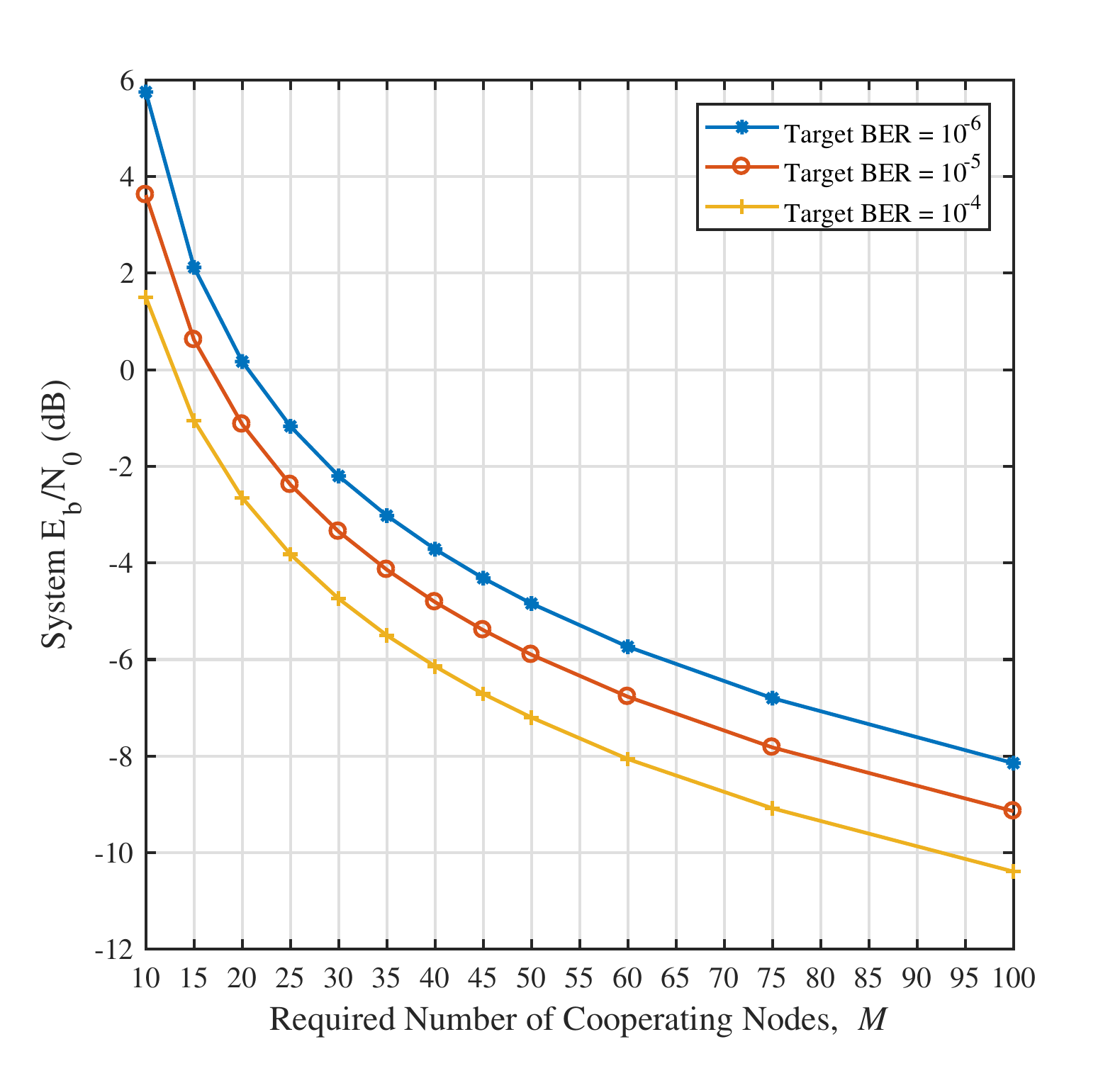}
	\caption{Required number of cooperating nodes to achieve a given target BER}
	\label{theory}
\end{figure}

Using the described system model, the performance of the uncoded DTB system is investigated for varying numbers of cooperating transmit nodes using Monte Carlo simulation. Perfect channel state information (CSI) and coherent detection at the receiver are assumed. The total transmit power and the energy per information bit, $E_{b}$, is kept the same across all DTB systems, irrespective to the number of transmit nodes used.  The results obtained for the uncoded DTB systems with Binary Phase Shift Keying (BPSK) modulation are shown in Figure~\ref{uncoded}.  These results show that the DTB system with 10 transmit nodes give a gain of about 41~dB over the SISO system at a Bit Error Rate (BER) of $10^{-5}$.  It should be noted that this gain includes both the spatial diversity and beamforming gains.  As a reference, a DTB system with $M$=10 nodes can give a beamforming gain of upto $10\log_{10}(M)$~=~10~dB.
In \cite{ishtiaq_18} we analysed the error rate performance of DTB and derived simple asymptotic (in the number of transmitters $M$) closed-form average BER expressions for multi-order Quadrature Amplitude Modulation (QAM) and Phase Shift Keying (PSK) modulation schemes.  Using the derived expressions, the BER ($P_{\text{E}}$) for the uncoded DTB with BPSK modulation in Independent Rayleigh fading with perfect synchronisation can be expressed as,
\begin{equation}\label{ishtiagBER}
P_{\text{E}} = Q \left(\sqrt{\frac{\pi M E_{b}/N_{0}}{2+(4-\pi)E_{b}/N_{0}}} \right)
\end{equation} 
where $Q(.)$ is the Gaussian $Q$-function \cite{simon_05}. Analytic results generated using Eq. (\ref{ishtiagBER}) are presented in Figure \ref{uncoded} and \ref{theory}.  \hspace{0.1cm} Figure \ref{theory} shows the minimum number of distributed nodes required to achieve a given BER and system $E_{b}/N_{0}$.
It can be observed from Figure \ref{theory} that the uncoded DTB system operating at an $E_{b}/N_{0}$ of -3~dB requires at least 28 transmit nodes to achieve a target BER of $10^{-5}$.  In the following sections, we describe and propose a scheme for constructing powerful short LDPC codes. We show these codes applied to the DTB system drastically improves its performance gains.  For example, a constructed 1/2-rate LDPC code of block-length 480 bits can reduce the number of required transmit nodes from 28 to 6 to achieve the previously indicated BER and $E_{b}/N_{0}$.

\hfill
\section{Construction of Short LDPC Codes}
\label{section_ldpc}
Low-density parity-check (LDPC) codes are a type of linear block error correction code with parity-check matrices (H) that contain a very small number of non-zero entries.  The sparseness of H guarantees the minimum Hamming distance and decoding complexity increase linearly with the code length.  LDPC codes were invented by Gallager in 1962 \cite{gallager_63}. Some of the published results show that these codes can perform 0.0045~dB away from the Shannon Limit \cite{chung_01}.   LDPC codes have been adopted in several standards including IEEE 802.16 (WiMAX) and DVB-S2 \cite{dtbs2_09}.

\begin{figure}[t]
        \centering\small
        \includegraphics[width=.45\textwidth]{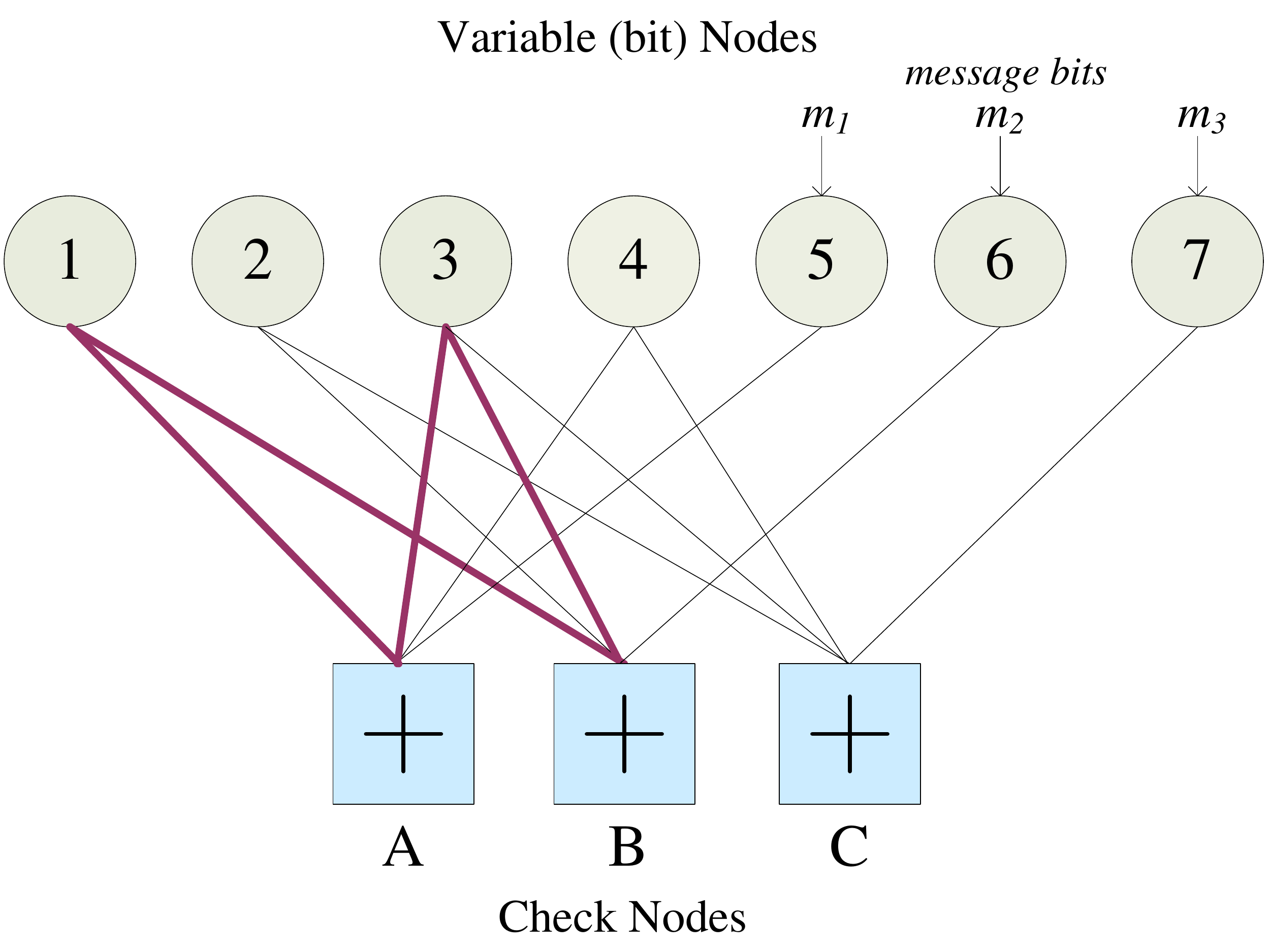}
        \caption{Tanner graph of the Hamming Code (7,4)}
        \label{hamming}
\end{figure}

An LDPC code can be denoted by $C(N, K)$, where $N$ and $K$ are the block-length and dimension (length of the message sequence) respectively.  The rate of $C$ can be expressed as $R = K/N$.  The number of parity bits is $(N-K)$.  Each parity bit corresponds to a parity-check equation.  All the $(N-K)$ parity-check equations can be represented by a parity-check matrix H.  All valid codewords in $C (\textbf{c} \epsilon C)$ satisfies the following equation,
\begin{equation}\label{syndro}
\text{H}\textbf{c}^{T} = 0
\end{equation} 
where $\textbf{c}$ is a codeword of block-length $N$ and H is the parity-check matrix with order $(N-K)\times N$.  Unlike the classical block codes (e.g. Reed-Solomon and BCH codes, which are usually decoded using algebraic decoding algorithms), LDPC codes are decoded iteratively using the graphical representation (Tanner graphs \cite{tanner_81}) of the parity-check matrix H of the code.  The iterative decoding algorithms operate alternatively on the bit nodes and check nodes (in the Tanner graph) to find the most likely codeword \textbf{c} that satisfies the condition H$\textbf{c}^{T}$=~0.   A Tanner graph can be denoted by $G(V,C,E)$ where $V$ and $C$ are the set of variable and check nodes respectively and $E$ is the set of edges.  
In the graph, each variable node represents a code bit, while each check node represents a parity check equation of the LDPC code.  The number of ones in the parity-check matrix H is equal to the number of edges in the Tanner graph.  As an example, the parity-check matrix for the Hamming block code (7,4) is given in (\ref{hammingH}) and corresponding Tanner graph is shown in Figure \ref{hamming}.
\begin{equation}\label{hammingH}
\text{H}=
  \left[ {\begin{array}{ccccccc}
   1&0&1&1&1&0&0 \\
   1&1&1&0&0&1&0 \\
   0&0&1&1&0&0&1 \\
  \end{array} } \right]
\end{equation}

The length of \textit{cycles} in the graph is an important performance metric of the LDPC code.  A cycle in a graph is a path defined by a sequence of connected vertices and edges where each edge is only visited once and the path starts and ends at the same vertex.  The number of edges in the cycle gives the length of the cycle.  A cycle of length 4 is highlighted in Figure \ref{hamming}.  The \textit{girth} of the LDPC code is defined by the cycle of minimum length.   Girth of the code is closely related to the minimum Hamming distance of the code, hence the error correcting capability of the code \cite{tanner_81}.  Short cycles in the Tanner graphs of LDPC codes degrade code performance as it affects the independence of the extrinsic information exchanged in the iterative decoding, preventing the iterative decoding algorithm from converging. Hence, LDPC codes with large girths are always preferred for maximising performance of the code.

\hfill
\subsection{Protograph-based LDPC Codes}
Encoding of LDPC codes is generally computationally intensive due to their large generator matrix sizes.  However, there are groups of LDPC codes with parity-check matrices specially designed to enable less computationally intensive encoding.  Examples of easily encodable LDPC codes include quasi-cyclic (QC) LDPC codes \cite{fosso_04} and repeat-accumulate (RA) LDPC codes \cite{yang_04}.  This paper focuses on designing QC-LDPC codes using protographs (P-LDPC).  A protograph or projected graph is a Tanner graph with a relatively smaller number of nodes \cite{thorpe_03}.  LDPC codes with large block-lengths can be constructed by generating multiple copies of a given protograph followed by permuting edges of the same type.  This copy-and-permute operation is also referred to as \textit{lifting} a protograph.  
A protograph $(n,k)$ lifted by an order of $v$ generates a code with block-length $vn$ and dimension $vk$.  It should be noted that the lifted code has exactly the same code-rate as the protograph.

\begin{figure}[b]
        \centering\small
        \includegraphics[width=.30\textwidth]{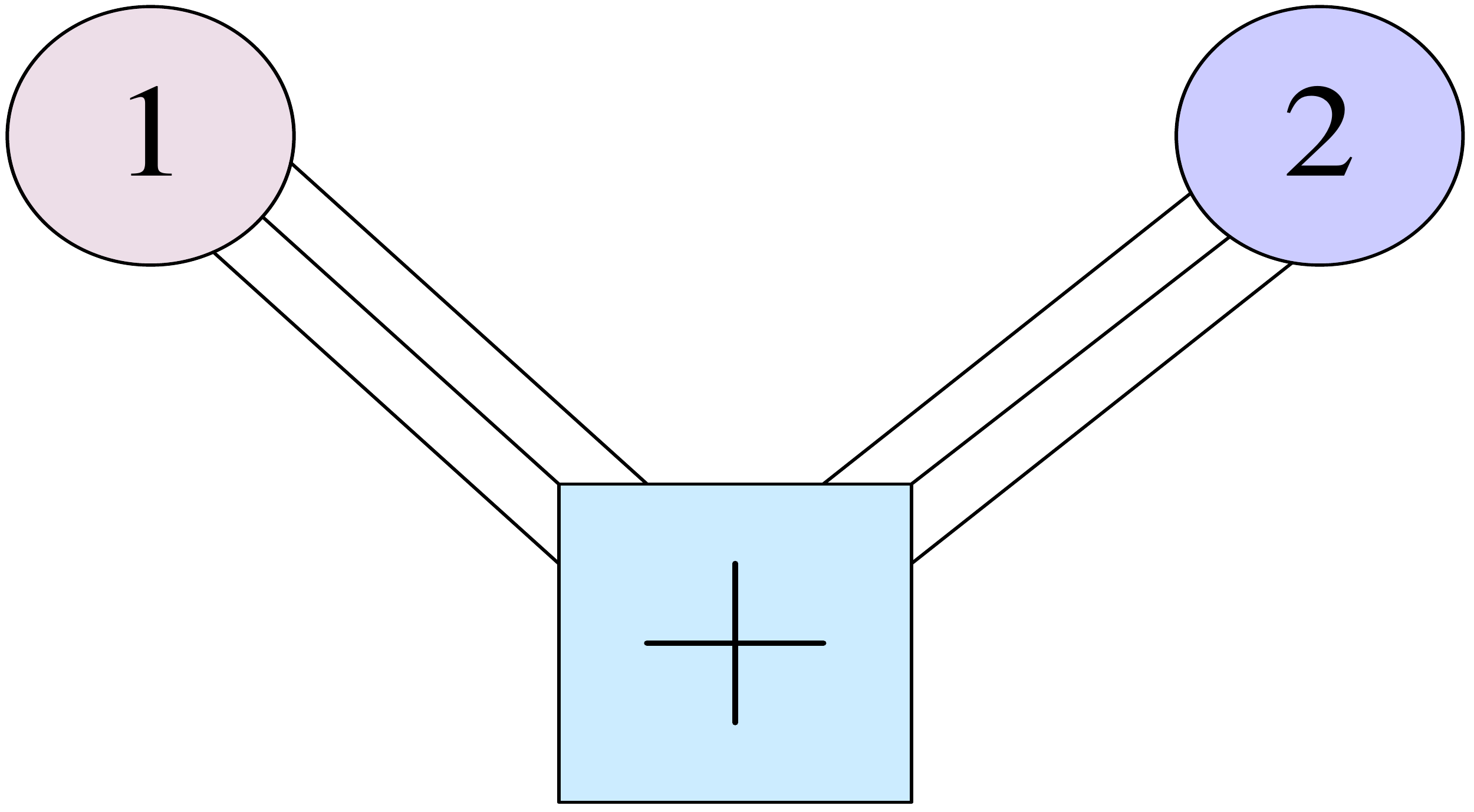}
        \caption{Multi-edged protograph}
        \label{multiEdgePro}
\end{figure}

\begin{figure}[b]
        \centering\small
        \includegraphics[width=.48\textwidth]{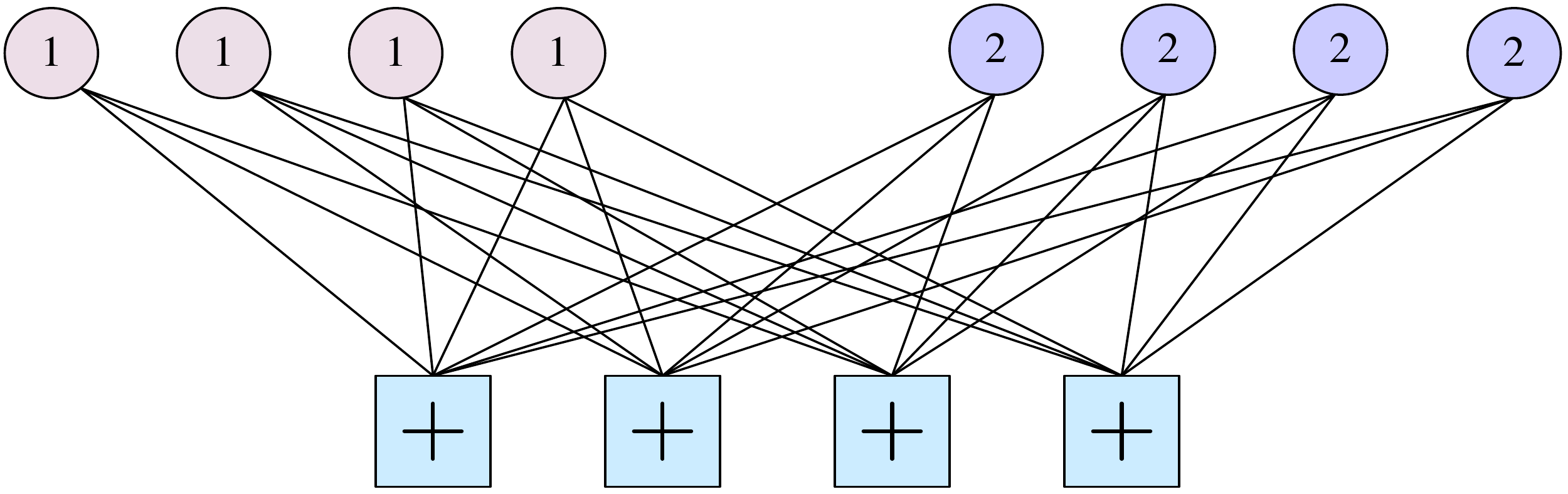}
        \caption{Lifting a multi-edged protograph}
        \label{liftingEg}
\end{figure}

An example of a multi-edged protograph with its parity-check matrix (protomatrix), $H_{p} = [3 \text{  } 3]$ is shown in Figure \ref{multiEdgePro}.  The resulting graph by lifting the protograph by a factor of 4 is shown in Figure \ref{liftingEg}.
The parity-check matrix of the lifted code is: 
\begin{equation}\label{multiH}
\text{H}=
  \left[ {\begin{array}{cccccccc}
1&0&1&1&1&1&0&1 \\
1&1&1&0&0&1&1&1 \\
1&0&1&1&1&1&0&1 \\
1&1&1&0&0&1&1&1 \\ 
  \end{array} } \right]
\end{equation}

\hfill
\subsection{Proposed Code Construction Scheme}
\begin{figure}[b!]
        \centering\small
        \includegraphics[width=.48\textwidth]{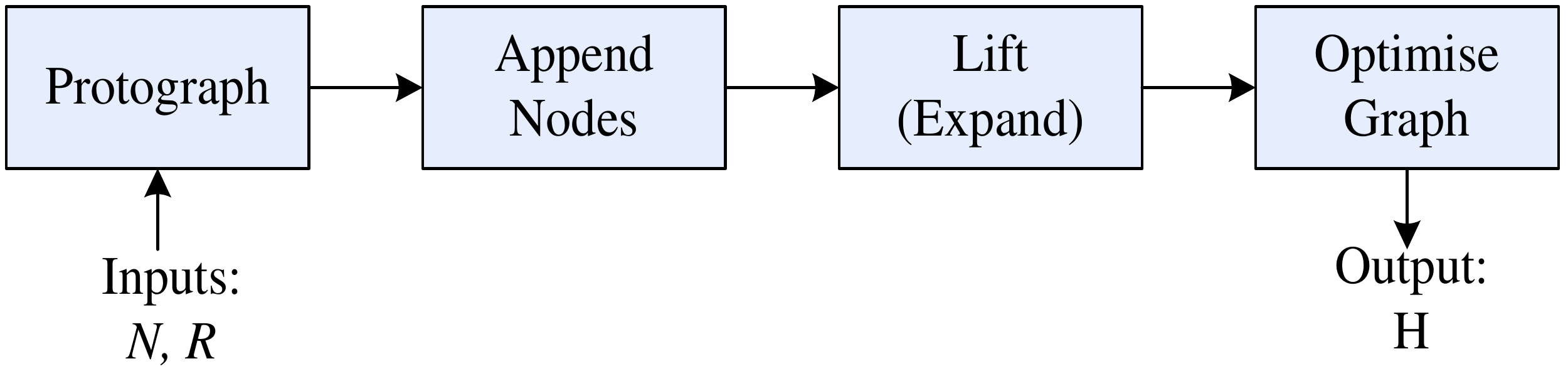}
        \caption{Proposed code construction scheme}
        \label{proposed}
\end{figure}

\begin{figure}[h!]
        \centering\small
        \includegraphics[width=.48\textwidth]{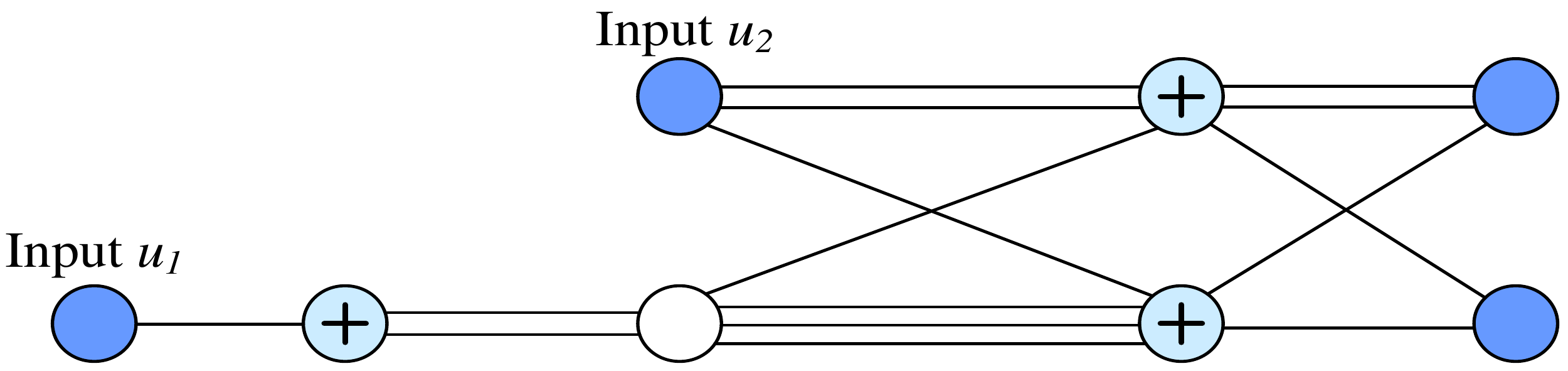}
        \caption{The protograph selected for code construction}
        \label{AR4JA}
\end{figure}

\begin{figure}[t]
        \centering\small
        \includegraphics[width=.48\textwidth]{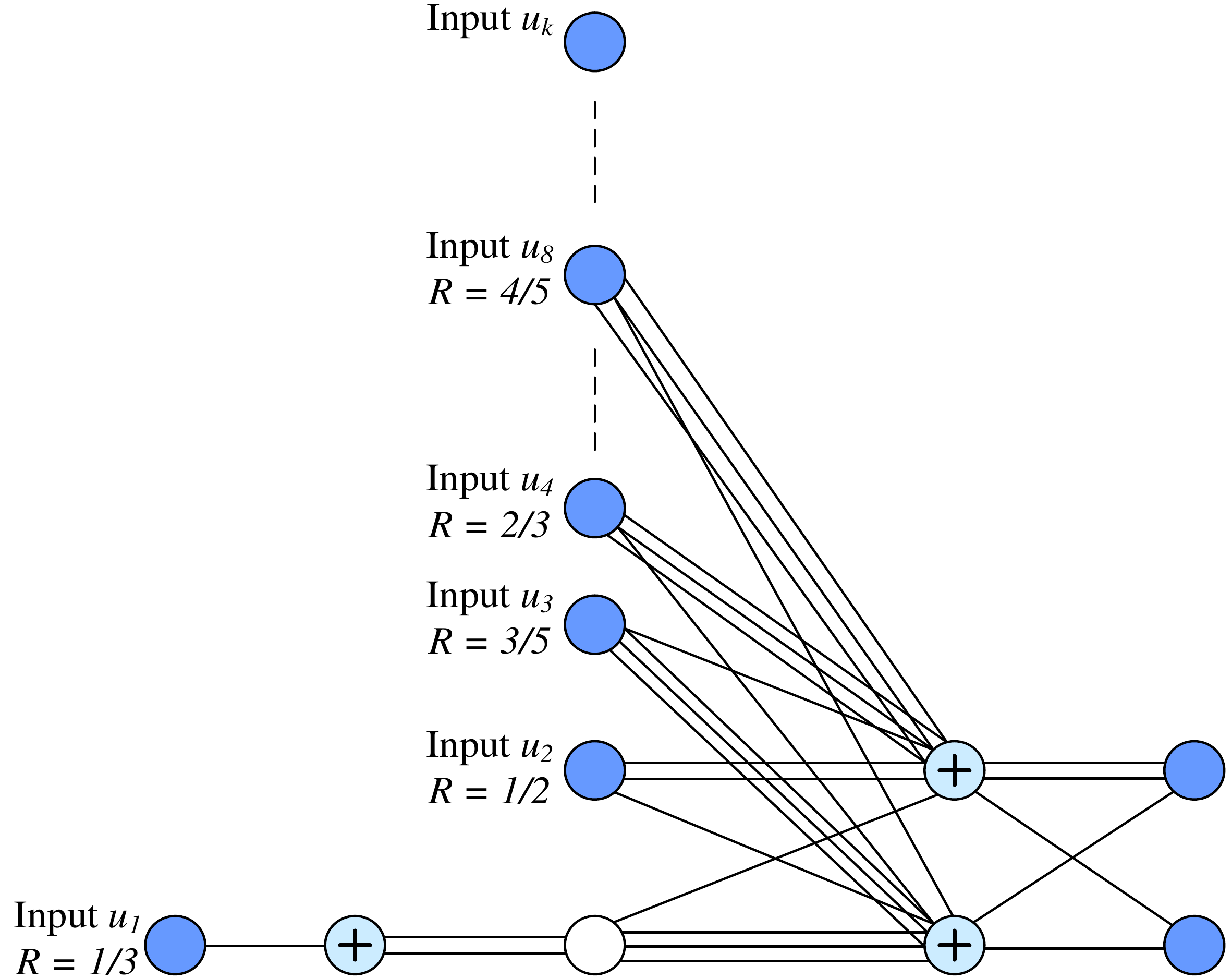}
        \caption{Node appending process}
        \label{AR4JA_append}
\end{figure}

The proposed code construction algorithm involves finding a parity-check matrix that maximises the girth of the lifted Tanner graph.  The problem of finding the optimal parity-check matrix is formulated as an optimisation problem and solved using the compact Genetic algorithm described in \cite{harik_99}.  The compact Genetic algorithm is used to search through the permutations to find the code with the largest girth. A computationally efficient algorithm for computing girth of a planar graph in linear time is proposed in \cite{chang_11}.  To reduce the computational intensity of the encoder and code construction process, quasi-cyclic LDPC codes are generated by limiting the search to only block circulants (i.e. cyclic permutations). The objective function for the formulated problem is an integer ‘multi-optima’ optimisation function.  One advantage of using this function is that it allows the generation of multiple non-identical parity-check matrices with large girths.  This feature can be used to improve the physical layer security of the system by varying the shared parity-check matrix at pre-defined time intervals.  Identical parity-check matrices can be generated at both the transmitter and receiver by varying some of the parameters used by the Genetic algorithm.  These parameters may include the seed of the pseudo-random generators, initialisation point of the search space, the step size or a bounded girth.  
The proposed code construction scheme is depicted in Figure \ref{proposed}.
The protograph used in the proposed construction method is shown in Figure \ref{AR4JA}.
This graph, commonly referred to as Accumulate Repeat-4 Jagged Accumulate (AR4JA), is one of the protographs proposed in \cite{divsa_09} by researchers at the Jet Propulsion Laboratory (JPL).  AR4JA is a 1/2 rate code with three check nodes and five variable nodes.  Out of the five variable nodes, one node (blank circle) is a punctured node which takes part in the encoding and decoding but is not transmitted. Protographs with other code-rates are obtained by appending additional nodes to the base protograph until the desired code-rate is achieved \cite{nguyen_13}.  The node appending process is depicted in Figure \ref{AR4JA_append}.  
Rate 1/3 is achieved by removing $u_{2}$ from the base protograph. For the protograph considered, the number of additional nodes $n_{R}$ required  to reach a target code-rate $R$ can be expressed as,
\begin{equation}\label{errorCap}
n_{R} =  \lfloor{\frac{(1-3R)}{(R-1)}}\rfloor  
\end{equation}
Once $n_{R}$ is determined, the lifting order $v$  required to achieve block-length ($N$) can be calculated using,
\begin{equation}\label{liftOrder}
v = \lfloor{\frac{N}{(n_{R}+3)}}\rfloor 
\end{equation}



A limitation of the proposed code construction algorithm is that the generated code-rates are restricted to certain values, e.g. 0.33, 0.5, 0.6, .., etc.  To overcome this issue, the proposed scheme is modified to allow generation of additional rates.  This modification involves truncating the lifted parity-check matrix and then optimising the truncated matrix using the compact Genetic algorithm.  This method gives more flexibility in generating codes with a much larger range of code-rates.  For example, it can be used to generate code-rates less than the minimum defined rate of 1/3. In addition, this modification can also be used to generate code-rates between any two defined code-rates, e.g. rates between 1/2 and 3/5.   

\hfill
\subsection{Performance of P-LDPC-coded SISO Systems}
\begin{figure}[h!]
	\centering\small
	\includegraphics[width=.53\textwidth]{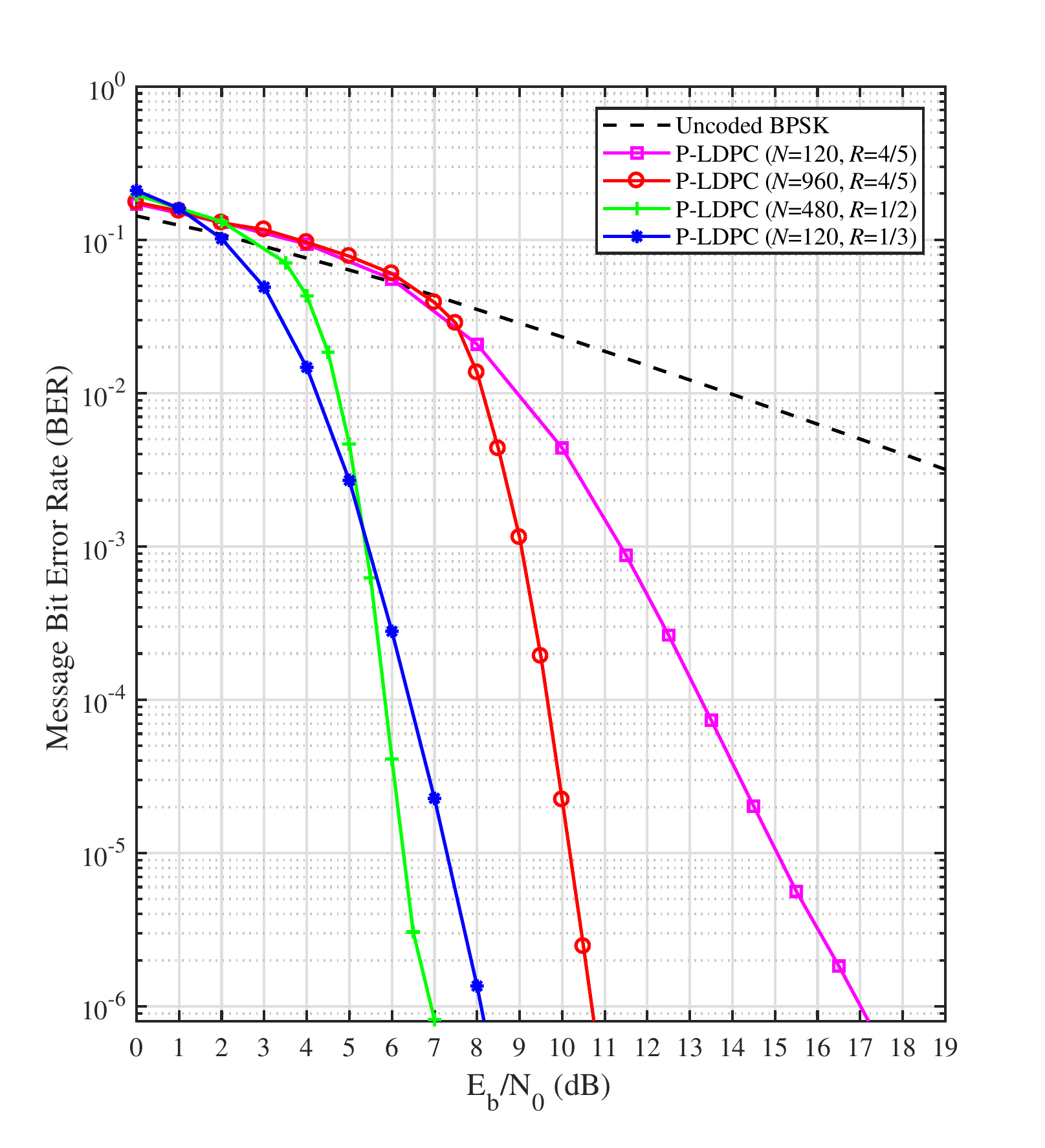}
	\caption{Performance of constructed P-LDPC codes in Independent Rayleigh fading}
	\label{codePerms}
\end{figure}

Performance of some of the codes constructed using the proposed scheme is shown in Figure \ref{codePerms}.  These codes are constructed for block-lengths of 960, 480 and 120 bits with code-rates of 4/5, 1/2 and 1/3. The belief propagation algorithm \cite{gallager_63} is used for soft-decision decoding of the P-LDPC codes. The performance is evaluated for the coded system with Binary Phase Shift Keying (BPSK) modulation in the Independent Rayleigh fading channel using Monte Carlo simulation.  The results generated are compared to the uncoded BPSK system at a BER of $10^{-6}$.  It has been shown in Figure \ref{uncoded} that the uncoded system attains this error-rate at an $E_{b}/N_{0}$ of 53.9~dB.   
The results presented in Figure \ref{codePerms} show that the codes ($N$=120, $R$=4/5), ($N$=960, $R$=4/5), ($N$=480, $R$=1/2) and ($N$=120, $R$=1/3) give coding gains of approximately 36.9, 43.2, 47.0 and 45.9~dB respectively over the uncoded system at the BER of $10^{-6}$.  Figure \ref{codePerms} also shows that the code ($N$=960, $R$=4/5) of block-length 960 bits performs 6.3~dB better than the shorter code ($N$=120, $R$=4/5) of block-length 120 bits for the same code-rate ($R$=4/5).

\hfill
\section{Performance of Coded-DTB Systems}
\label{section_codedDTB}
\begin{figure}[h!]
	\centering\small
	\includegraphics[width=.53\textwidth]{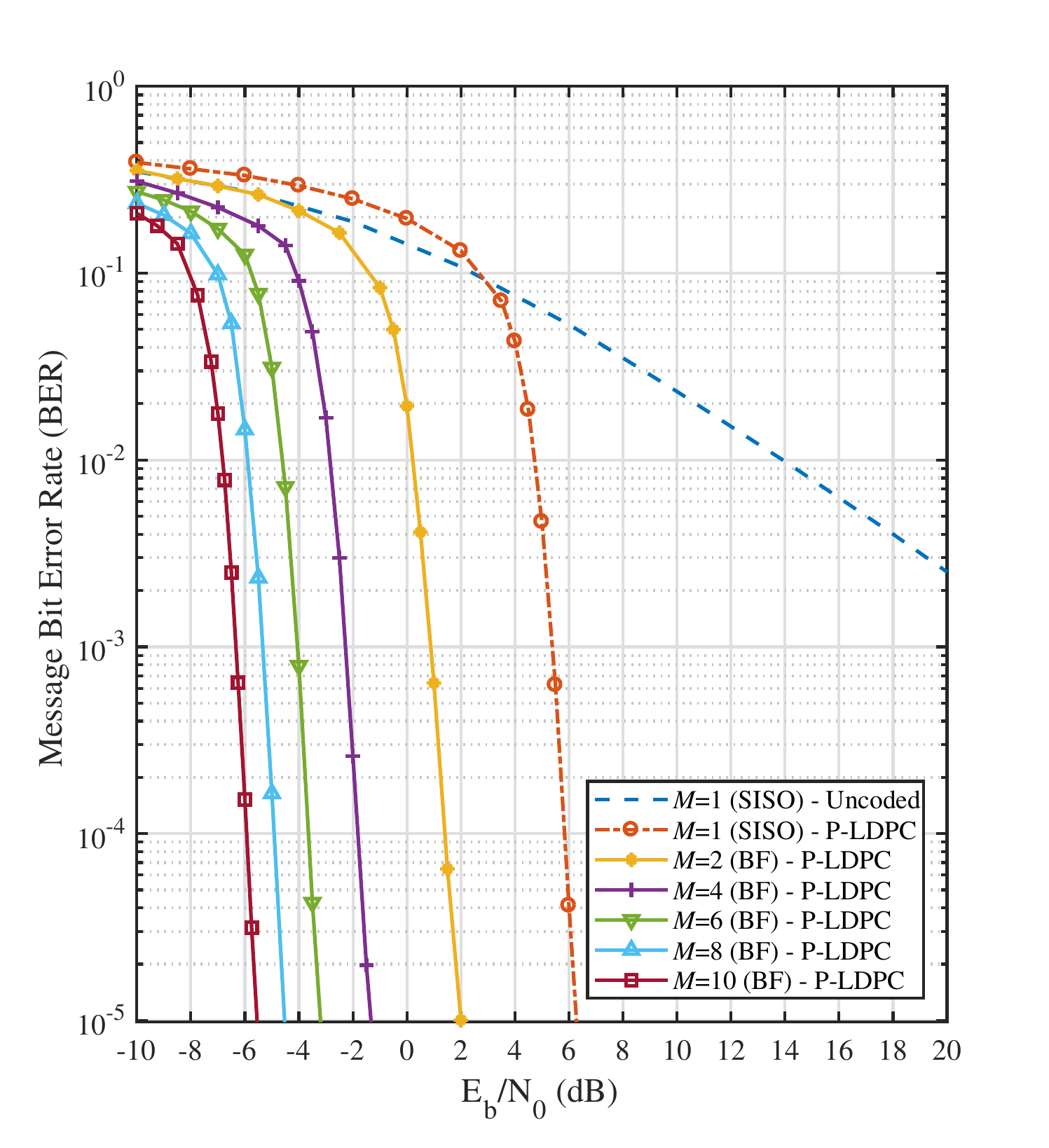}
	\caption{Performance of P-LDPC(480, 240)-coded DTB for varying number of cooperating nodes in Independent Rayleigh fading}
	\label{ldpc480}
\end{figure}

In this section, we investigate performance of P-LDPC-coded DTB systems and compare their performance with other reference coded DTB systems.  The performance is evaluated for the Independent Rayleigh fading channel.  Figure \ref{ldpc480} shows performance of a P-LDPC(480, 240)-coded DTB system with different numbers of cooperating transmit nodes ($M$).  The P-LDPC(480, 240) is a 1/2-rate code constructed using the proposed scheme.  The construction uses the base protograph with no additional nodes added.  The lifting factor used is 120, resulting in block and message lengths of 480 and 240 bits respectively.  The soft log-likelihood ratios (LLR) for the decoder are computed using the approximation given by \cite{hou_01},
\begin{equation}\label{LLR}
\text{LLR} =  \frac{2}{\sigma^{2}_{n}}r.a,
\end{equation}
where $a$ is the channel fading gain (known to the receiver), $r$ is the received signal and  $\sigma^{2}_{n}$ is the variance of the additive white Gaussian noise $n \sim \mathcal{N}(0,\,\sigma^{2}_{n})$. Key performance parameters from Figure \ref{ldpc480} are tabulated in Table \ref{codedDTBtable}.  The performance results presented in this table are all in dBs.
\begin{table}[]
	\caption{Performance of P-LDPC-coded DTB systems (@ BER of $10^{-5}$)}
	\label{codedDTBtable}
	\centering
	\begin{tabular}{c|cc|cc}
		\hline
		Number    & \multicolumn{2}{c|}{$E_{b}/N_{0}$ gain over u-SISO}  & \textbf{Coding} & \textbf{Beamforming} \\  \cline{2-3}
		of nodes    & \textbf{Coded} & \textbf{Uncoded}  & \textbf{gain} &  \textbf{gain}\\
		 M     & \textbf{DTB} & \textbf{DTB}  &  & (Theoretical) \\
		\hline
		2     & 42.0     & 22.1   & 19.9    &   3.0\\
		4     & 45.3     & 33.1   & 12.2    &   6.0\\
		6     & 47.2     & 37.1   & 10.1    &  7.8\\
		8     & 48.5     & 39.5   & 9.0     &  9.0\\
		10     & 49.5     & 41.1  & 8.4    &  10.0\\
		\hline 
	\end{tabular}
\end{table}
The gains shown in the column titled \textit{Coded DTB} includes combined beamforming, spatial diversity and coding gains over the uncoded non-beamforming single-antenna SISO system.  Similarly the figures shown in the column titled \textit{Uncoded DTB} gives the combined beamforming and spatial diversity gains.  The performance results for the uncoded DTB have been discussed in detail in Section \ref{section_DTB}.  These results show that the uncoded DTB system operating at an $E_{b}/N_{0}$ of -3~dB requires at least 28 transmit nodes to achieve a target BER of $10^{-5}$.  Compared with this, the results presented in Figure \ref{ldpc480} for the coded DTB system show that using a 1/2-rate P-LDPC code of block-length 480 bits requires only 6 transmit nodes to achieve the same BER and $E_{b}/N_{0}$.
\begin{figure}[h!]
	\centering\small
	\includegraphics[width=.53\textwidth]{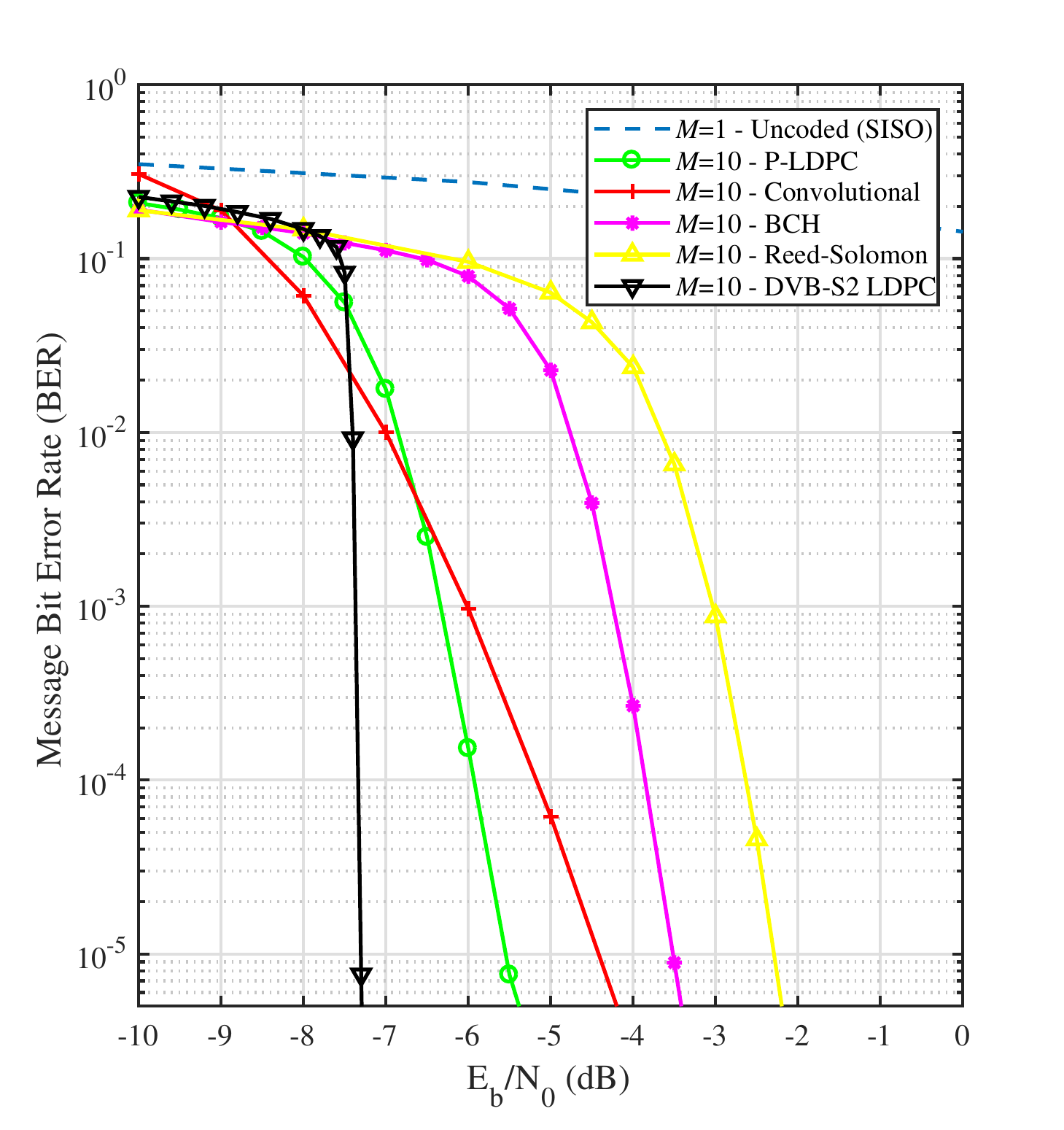}
	\caption{Performance comparison P-LDPC($N$=480, $R$=1/2)-coded DTB with other coded-systems}
	\label{allCodes}
\end{figure}

The performance of the P-LDPC(480, 240)-coded DTB system is also compared with other reference coded-DTB systems.  This includes DTB systems with Bose-Chaudhuri-Hocquenghem (BCH), Reed-Solomon and Convolutional codes.  All these codes are of rate and length 1/2 and 480 bits respectively.  In addition, we also compared its performance with the LDPC code adopted in the DVB-S2 standard \cite{dtbs2_09} which has a block-length of 64800 bits and rate 1/2.  The performance results obtained are presented in Figure~\ref{allCodes}.  More details of the codes and the soft-decision decoding (SDD) algorithms used to assess their performance are shown in Table~\ref{sddAlgos}.     
\begin{table}[t]
	\caption{Parameters of the coded-DTB systems}
	\label{sddAlgos}
	\centering
	\begin{tabular}{l|l|l}
		\hline
		\textbf{Coding}    & \textbf{Construction} & \textbf{SDD algorithm}   \\
		\hline
		BCH             & Shortening mother code,      & Chase type-II \cite{chase_72},   \\
						&  (511, 259)  & Reduced set ($\sim$ 1000) \\
						\hline
		Reed-Solomon     & Shortening mother code,      & Chase type-II \cite{chase_72}, \\
						&   (255, 223, GF(256))  &     Reduced set ($\sim$ 1000)   \\		
						\hline				
		Convolutional   &  Constraint length 7,     & Soft-Viterbi \cite{viterbi_67},   \\
						&	Polynomial [133 171]	&  Traceback length 56 \\
						\hline
		LDPC (DVB-S2)   &   DVB-S2 \cite{dtbs2_09}      &   Belief propagation \cite{gallager_63}\\
		\hline
		P-LDPC     & Proposed algorithm     &    Belief propagation \cite{gallager_63} \\
		\hline 
	\end{tabular}
\end{table}
The presented results show that P-LDPC(480,240)-coded DTB system with 10 transmit nodes give gains of approximately 1.1~dB, 2.0~dB and 3.3~dB over the corresponding systems with Convolutional, BCH and Reed-Solomon codes respectively, at a BER of $10^{-5}$.  It is also observed that the LDPC(DVB-S2)-coded system with a block-length of 64800 bits performs $\sim$1.8~dB better than the P-LDPC(480,240)-coded system (at the same BER).  This is expected as the block-length of the code used in the LDPC(DVB-S2) system is 135 times longer than the block-length used for the P-LDPC system.

\hfill
\section{Conclusion}
\label{section_conclu}
This paper evaluated performance of the uncoded DTB system and proposed a robust adaptive coding scheme. 
The proposed scheme constructs random quasi-cyclic LDPC codes using protographs and optimises them using a compact Genetic algorithm.  The performance of some of the codes constructed using the proposed scheme was investigated in the Independent Rayleigh fading channel.  The performance results were compared with the uncoded and other coded-DTB systems.  The results obtained show large gains over the compared systems.  It was observed that the proposed algorithm was very flexible in generating powerful LDPC codes with varying code-rates and block-lengths.  This adaptive feature is very useful to optimise system resources and provide reliable communication under a large variation of channel conditions using a smaller number of distributed transmit nodes.

\hfill
\section*{Acknowledgment}
The work presented in this paper is conducted as part of the research project titled \textit{Resilient Communication - Distributed Transmit Beamforming}.  This project is supported by the Next Generation Technologies Fund managed by Defence Science and Technology (DST) Group.  A simplified proof-of-concept waveform to verify the feasibility of DTB has now been implemented and successfully tested outdoors. Current status and details of ongoing activities of the project are discussed in \cite{leak_18}.  

The authors would like to acknowledge Dr Gerald Bolding (DST Group) for the valuable suggestions made to improve the P-LDPC code construction scheme described in this paper.   

\hfill

\end{document}